**Understanding model behavior using loops that matter.**

William Schoenberg[1,2], Pål Davidsen[1], Robert Eberlein[2]

[1] University of Bergen, Norway
System Dynamics Group, Department of Geography,
University of Bergen, Fosswinckels gt. 6,
Postboks 7802, N-5020 Bergen, Norway

[2] isee systems inc. Lebanon, New Hampshire USA
Wheelock Office Park
31 Old Etna Rd, Suite 7N
Lebanon, NH 03766 USA

## Abstract

The relationship between structure and behavior is central to System Dynamics, but effective tools required to understand that relationship still elude us. The current state of the art in the field of loop dominance analysis relies on either practitioner intuition and experience or complex algorithmic manipulation in the form of eigenvalue analysis or pathway participation metrics. This paper presents a new and distinct numeric method based on a different measure, the loop score, to determine the contribution of a loop to a model's behavior at each instant in time. This allows us to discover the origin of model behavior. The method was inspired by observations of the patterns in the changes of the values of variables during simulations and has been tested and refined using empirical evaluation on a variety of models. The method also offers a promising approach to the visualization and aggregation of simulation results.

## The problem

The strong relationship between structure and behavior is fundamental to system dynamics (Sterman, 2000). A model uses parametric and structural assumptions to produce behavior. A system dynamics practitioner must generally engage in the following process: create a structure underlying the problem under consideration, understand how that structure works to produce that problem, and figure out how to improve that structure so as to address the problem. The second step in this process is the focus of this paper and is the key to successfully performing the third step. By referring back to the model structure, the practitioner can explain the reasons why the observed behavior has been generated (Richardson, 1996). Based on that understanding, the practitioner may propose changes in input values or model structure (system configuration) that will cause more favorable behavior to be produced.

The current state of the art in the field relies on either practitioner intuition and experience (the craft of modeling and model analysis) or complex algorithmic analysis. The former is taught as part of the methodology of model building, while the latter comes from 40 years of work on techniques to derive and explain model behavior based on the analysis of structure (see for example: Graham, 1977; Forrester 1982; Eberlein, 1984; Davidsen, 1991; Mojtahedzadeh, 1996; Ford, 1999; Saleh, 2002; Mojtahedzadeh et al., 2004; Güneralp, 2006; Gonçalves, 2009; Saleh et al., 2010, Kampmann, 2012; Hayward and Boswell, 2014; Moxnes and Davidsen, 2016; Oliva, 2016; Sato, 2016; and Hayward and Roach, 2017).

Ford (1999, p.4-5) clearly stated the needs of the system dynamics field as they apply to loop dominance analysis:

> *"To rigorously analyze loop dominance in all but small and simple models and effectively apply analysis results, system dynamicists need at least two things: (1) automated analysis tools applicable to models with many loops and (2) a clear and unambiguous understanding of loop dominance and how it impacts system behavior."*

In this paper, we discuss a new loop dominance analysis method and demonstrate its performance on the two tests set out by Ford.

We define loop dominance as a concept which relates to the entirety of a model, as opposed to loop dominance being something that affects a single stock. For loop dominance to apply to the entire model, we require that all stocks are connected to each other by the network of feedback loops in the model. For models where there are stocks that do not share feedback loops, we consider each subcomponent of inter-related feedback loops individually, and we refer to each model substructure as having a separate loop dominance profile. Our measurement of loop dominance is specific to the particular time period selected for analysis. We say that a loop (or set of loops) is dominant if the loop(s) describe at least 50% of the observed change in behavior across all stocks in the model over the selected time period.

The purpose of this research is to create a loop dominance analysis method that is accurate and practical. While this new method may not generate new insights relative to existing analysis methods, it has two characteristics that make it more practical. First, it is relatively easy to understand for people who have built and analyzed system dynamics models, even those with a limited mathematical background. Second, it lends itself completely to presenting behavior over time graphs that mix loop dominance measures and model behavior, making it easy to see the evolution of loop dominance. These characteristics, combined with the very straightforward computational techniques we employ, mean that such an analysis can easily be built into existing system dynamics modeling environments. This will make finding the loops that matter over time as easy as it is to watch a graph of behavior unfold over time. Such ease of use means the methods will routinely remove one more obstacle in our ability as a field to do good work.

First, we discuss the existing approaches used to automate the process of discovering loop dominance. Thereafter, we introduce our new method, starting with links and building up to loops. Then we demonstrate the new method applied to three models that have already been analyzed using existing automated loop dominance analysis techniques, so as to demonstrate how well the new approach works. Finally, we conclude with a discussion of the benefits and weaknesses of the new method.

## Literature review

The current state of the art in the use of mathematical methods for determining loop dominance revolves around two methods. The first one is based on eigenvalue elasticity analysis, the second one, uses the pathway participation metric and causal pathways.

**Eigenvalues and eigenvectors, specifically eigenvalue elasticity analysis (EEA)**

Forrester (1982) was the first to document that eigenvalue elasticities could be used to explain the relative contributions of different loops in models of linear systems. Since then, the formal method of eigenvalue elasticity analysis (EEA) has been further developed and is now used to determine how model structure produces the dynamic modes of behavior in a model, specifically those modes of behavior characterizing the state variables or stocks, (Saleh, 2002), (Kampmann et al., 2006), (Saleh et al., 2010), (Oliva, 2016). Using EEA, the combination of behavior modes expressed by a model is characterized by the eigenvalues and eigenvectors of that model. This analysis is built on the observation that the behavior of a linear system can be expressed as a weighted combination of behavior modes, each characterized by a decoupled (or pairwise coupled) set of eigenvalues (Saleh et. al., 2010). EEA is applied to examine both link and loop significance with regard to the dynamic behavior of the model. It does so by identifying the relationship, expressed in the form of the elasticity, between the parameters that make up the gains of individual feedback loops in the model and the eigenvalues (and sometimes eigenvectors) that characterizes the dynamic behavior of the model. The significance of a loop to generating behavior is expressed by the eigenvalue elasticity of its gain (how strongly a change in the loop gain impacts the eigenvalue associated with the behavior of interest). Note that this may not only be used to identify the root cause of a model´s behavior, but also the leverage points for controlling a system (policy entry points) provided the model is an accurate representation of that system.

Kampmann (2012) developed the concept of the independent loop set (ILS) which maps all of the loops in the model onto a singular set of independent loops that produces the full behavior of the model so that the analysis can be effectively completed and interpreted. Oliva (2004) extended Kampmann's work on the ILS by developing an ILS composed only of geodetic loops which he termed the shortest independent loop set (SILS). That is the de-facto standard for determining which loops to analyze in EEA.

A disadvantage of the EEA methods is that they are mathematically complex and require a deep understanding of linear algebra. There are a variety of other limitations of the EEA method as implemented today (Naumov & Oliva, 2018) that are rooted in the specifics of the implementation itself, but not the general theory. The most important limitation of the current implementation of EEA is the requirement that models be made continuously differentiable, which is difficult for many models in use. An example of this shortcoming in the implementation is presented in Oliva (2016). When Oliva (2016) analyzed his service quality model, he had to, among other changes, remove a stock in order to produce a full rank system matrix, and had to change model equations to ensure the model was continuously differentiable. Those changes did have an impact on the simulation results, - as will such changes in general.

**Pathway participation metric (PPM) and other causal pathway techniques**

The pathway participation metric (PPM) approach does not use eigenvalues to describe model structure. Rather, it focuses on the links between variables (Mojtahedzadeh et al, 2004). The starting point of the PPM approach is the behavior of a single variable, typically a stock. The behavior of that single variable is partitioned in time, based on periods in which the variable maintains its slope and convexity (i.e. the first and second time derivatives do not change sign) (Mojtahedzadeh et al, 2004). The behavior of the variable in each of these phases is characterized by 7 patterns enumerated by Mojtahedzadeh et al, (2004). The PPM approach then determines dominance by tracing along the causal pathways between the stock under study and its ancestor stocks so as to determine which structure is the most influential one in terms of explaining the pattern of behavior exhibited by that stock during the selected phase. Mojtahedzadeh et al., (2004) explains that it does so by determining the magnitude of the change in the net flow of the stock under study by making minute changes to that stock. The method then compares these changes in the net flow in order to determine the change of the largest magnitude in the same direction as the stock under study, - thereby identifying the most important (dominant) pathway governing the behavior of that stock during that phase.

Relative to the very general EEA approach that yields results that are interpretable across the entire model, PPM is considerably more specific, only explaining the impact of feedback loops on specific stocks of interest. In addition, PPM does not identify the dominance of behavior modes and, consequently, only provides information about behavior that can be readily observed in the trajectory of stock behavior.

The advantage of the PPM approach, relative to the EEA approach, is that it does not require any manipulation of the model structure and that it's implementation does not require model structure to be continuous. Moreover, according to the research by Mojtahedzadeh (1996), the application of the PPM method will cause a convergence on a unique piece of structure as the one most influential with regard to the phase of its behavior under study. Kampmann and Oliva (2009) state that one of the key benefits to the PPM method is its ability to associate directly model behavior to structure.

Kampmann and Oliva (2009) have criticized PPM for its inability to clearly explain oscillatory behavior and also because PPM may fail to identify structure when there are two pathways of similar importance (Kampmann and Oliva, 2009). Hayward and Boswell (2014) have responded to those criticisms by simplifying PPM into the Loop Impact method. The Loop Impact method can be implemented in a standard system dynamics models (and software) by adding equations to the model. No change in the underlying software is required. The key difference of the Loop Impact method as compared to PPM is that it does not identify dominant pathways (impacts from one stock to another), but instead focuses on the direct impact that one stock has on another to identify loops which dominate the behavior of the selected stock. Pathways are chained together according to the structure of the model and these chains of pathways are used to measure the Loop Impact metric which yields insight into which loop dominate the behavior of the selected stock (Hayward and Boswell, 2014). In addition, the Loop Impact method identifies instances where multiple loops are required to explain the behavior of a stock.

Expanding on the work done by Hayward and Boswell (2014), Sato (2016) has modified the Loop Impact method specifically codifying the impacts of force in an engineering sense. Hayward and Roach (2017) have also developed a framework around the Loop Impact method couched in the mathematics of Newtonian physics, to explain the model as a series of interacting forces. The stated purpose of the underlying common research thread between these authors of the Loop Impact method and its derivatives, is to provide a more intuitive and complete understanding of loop dominance in system dynamics models.

## The loops that matter method

### Introduction to loops that matter

In this paper, we present the LTM (Loops That Matter) method which, under the categorization scheme of Duggan and Oliva (2013), algorithmically performs a "formal assessment of dominant structure and behavior" for models of any size, complexity, or dimensionality. Like PPM, this method is derived, in part, from observations of the way that experienced modelers perform analysis to determine the sources of observed behavior. Unlike EEA, and similar to PPM, this method does all of its calculations directly on the original model equations, walking the causal pathways between stocks through all intermediate variables making it easier to understand the measurements of loop contribution to model behavior and link contribution to feedback loop dominance. The method uses only values computed during a regular simulation of any model, - including models with discrete characteristics. Because there is no model transformation taking place, there is also no canonical form that definitively and uniquely defines the characteristics of model behavior the way that the eigenvalues and eigenvectors do in the case of linear models. Instead, we have developed a metric that tracks the concept of link gain such that the chain rule of differentiation can be used in defining a second metric we have developed to track the concept of loop gain. This guarantees that the description of dynamic behavior will be the same, independent of the exact form of structurally equivalent models (many variables with simple equations or few variable with complex equations). This property of the LTM approach is proven in Appendix II. While our definition of loop dominance is not demonstrably unique among structurally equivalent models, it seems unlikely that alternative sets of equations would change the ranking of loop scores so as to give substantively different results.

LTM is applicable to more than regular continuous SD models. LTM is applicable to agent based as well as discrete time models as long as the structure of these model is represented as a network of equations to be evaluated at known time points. The reason that LTM is so broadly applicable is because LTM is only computed at each time point that the model is calculated, using only the existing causal relationships and variable value computations as inputs. LTM ensures that it does not affect the validity of the analyzed model by not manipulating equations or using variable-values not computed as part of the regular simulation. An example of a manipulation, required by a method, that affects the validity of a model under analysis is a change a discrete, integer only variable, say something like product color from 2 to 2.1. This change would result in a logical error in the model (even though the model may be calculated without error) because the modeler did not anticipate a non-integer value for that variable and, therefore, the model responds in an unanticipated way.

LTM introduces two metrics. The first one, the link score, is used to measure the contribution and polarity of a link between an independent variable and a dependent variable. The second one is the loop score that measures the contribution of a feedback loop to the behavior of the model and is indicative of the feedback polarity. Loop scores and link scores are calculated at each time interval during the simulation run. The analysis of the relative loop scores at a particular point in time identifies the loops that dominate that model behavior. The display of the relative loop scores over time contributes to our understanding of why the model behaves the way it does, i.e. points to the feedback loops that govern current behavior. By definition, loop scores are completely insensitive to the number of variables and links in a loop.

We use the standard definition of a loop: It is a set of interconnections between variables in a model that form a closed path from a variable back to itself. By definition any such loop must include at least one state variable (so as to avoid a simultaneity). We refer to the interconnections as links. Loop scores are computed as products of link scores, and the definition of a link score is tailored to this specific use. The link score computation has been defined for the sole purpose of determining loop dominance, specifically to be used in the loop score calculations.

**Introduction to the link score measure**

The link score measures the contribution of a value change in an independent variable to a value change in a dependent variable and also the associated polarity. A link score is calculated for each link in the network of model equations, - including the links from flows to stocks. Because links from flows to stocks constitute an integration process, whereas links into other variables do not, we have devised two methods for calculating the link score, where both have the same conceptual interpretation. The first method we discuss below is for links which do not constitute an integration process, for instance a link from one auxiliary to another. The second method we discuss handles the links from flows to stocks that are integrations. Both methods produce a link score, and these link scores are directly comparable and are multiplied together to produce loop scores.

The link score is not a general metric that can be used to describe the contribution or importance of any specific link in isolation. It must be considered in the context of the loops it is contained in. The most obvious manifestation of the link score's lack of generality is that a link originating from a variable that is not changing its value (including a parameter), have a score that, by definition, is 0. This is so, because when a variable in a loop does not change value, the loop is inactive and, therefore, currently of no consequence. Even though the link score for all links originating from variables whose value do not change, are 0, such variables are still significant to the loop dominance analysis because they may, by serving as parameters in other equations, influence the way that other variables change. For example, a time constant might condition some other link score, which consequently changes the score of a loop containing that link.

**Defining link scores for links without integration**

To simplify the presentation, we will define the link score assuming there are two inputs (x and y ) to the dependent variable z characterized by the equation $z = f(x, y)$. This easily generalizes to the case where there are more (or fewer) inputs into the equation defining $z$.

The link score for the link x → z is computed using the formula in Equation 1.

*Equation 1: The discrete form for the link score equation which matches the implementation of the calculation and is computed each dt.*

$$LS(x \to z) = \begin{cases} \left( \left| \frac{\Delta_x z}{\Delta \mathrm{z}} \right| \cdot sign\left( \frac{\Delta_x z}{\Delta \mathrm{x}} \right) \right), \\ 0, \qquad \Delta \mathrm{z} = 0 \text{ or } \Delta \mathrm{x} = 0 \end{cases}$$

In Equation 1, Δz is the change in (the value of) z from the previous time to the current time. Δx is the change in $x$ over that interval. $\Delta_x z$ is the change in $z$ with respect to $x$ over that interval. From a computational perspective $\Delta_x z$ which we call the partial change in $z$ with respect to $x$, is the amount $z$ would have changed, conditionally, if $x$ had changed the amount it did, but $y$ had not changed (i.e. ceterus paribus). The first major term in Equation 1 represents the magnitude of the link score, the second one is the link score polarity.

The exceptions for no change in $x$ or $z$ are included for completeness. If x does not change then the links from x, and any loops involving x, will have a score of 0. (Similarly, if z does not change then any links into or out from z will have a score of 0 and any loops involving those links are inactive). In such cases, we simply define the value of the link score from x to z to be 0.

The first major term in Equation 1 is the link score magnitude $\left|\frac{\Delta_x z}{\Delta \mathrm{z}}\right|$ which describes the effect (force is a good analogy) that an input (i.e. independent variable) $x$ has on an output (i.e. dependent variable) $z$, relative to the total effect exerted on $z$. Unlike a partial derivative, which describes how sensitive $z$ is to changes in x, this magnitude describes how much the change in x has contributed to the total change in z.

The first term is dimensionless and represents the contribution of $x$ to the change in $z$. If all of the link scores have the same sign, it expresses the fraction of the change in $z$ that results from the change in $x$. If the equation for $z$ is linear (uses only addition and subtraction), then the values will always be in the range between 0 and 1. When there are both positive and negative link polarities in a nonlinear equation, the link score magnitude for such a nonlinear equation may take on a very large value. But this does not jeopardize the overall analysis of loop dominance because it is the relative values of loop contribution that are analyzed at each point in time and large magnitudes counteract each other in such a comparison.

The second major term in Equation 1 is the polarity of the link $sign\left(\frac{\Delta_x z}{\Delta \mathrm{x}}\right)$ which is defined as the sign of the partial difference at time $t$. This formulation is functionally the same as the one used in Richardson 1995. We use the partial difference notation in order to maintain consistency with the link score magnitude whereas Richardson uses the partial derivative notation. Our reformulation of Richardson's polarity makes it easier to calculate the link score because the $\Delta_x z$ value can be used in both the magnitude and polarity computation.

In Appendix I we show how Equation 1 can be recast using partial differences, and, in the limit, partial derivatives. This is helpful when comparing our approach with other metrics such as PPM (Mojtahedzadeh, 1996) and Loop Impact (Hayward and Boswell, 2014).

**Defining link scores for links which contain integration processes**

The link from flows to stocks represent an integration process, and stocks can only change over time as a result of the specific values of the associated flows. So time must pass in order for a flow rate value to materialize in a change in associated the stock value. This is very different from above where a dependent auxiliary changes *immediately* in response to a *change* in one of its independent variables. As an example, if a stock only has a single flow, and that flow value is a constant, that stock will change over the course of the simulation. Consequently, the link between the flow and the stock is active and the link score is non-zero. Since links from flows to stocks are the only links with this unique property, the computation is different in form, but constitutes the same concept as the one used in Equation 1.

Assume the stock equation $s = \int(i - o)$ where $s$ is the stock, $i$ is the inflow, and $o$ is the outflow. So we assume a single inflow and outflow for simplicity of presentation, - the generalization to multiple inflows and outflows is straightforward.

*Equation 2: Link score for all links from flows to stocks (both inflows and outflows are covered)*

$$Inflow: LS(i \rightarrow s) = \left(\left|\frac{i}{i-o}\right| * 1\right) \quad Outflow: LS(o \rightarrow s) = \left(\left|\frac{o}{i-o}\right| * -1\right)$$

If you compare Equation 2 to Equation 1 you will notice that they represent the same concept. The value of the flow ($i$ or $o$) is exactly the same concept as the partial change in $z$ with respect to $x$ ($\Delta_x z$) from Equation 1, where the flow is $x$ and the stock is $z$. From the perspective of the stock, - the value of the flow (when multiplied by dt) is the amount the stock will change if no other flows were active. The denominator ($i - o$) is the same concept as Δz: It is the change in the stock ($z$) from the previous time to the current time (again when multiplied by dt). Finally $sign\left(\frac{\Delta_x z}{\Delta \mathrm{x}}\right)$ is replaced with simply +1 or -1 based on whether the flow is an inflow or an out-flow, since the polarity of flows are fixed. These are the design elements that unite the two methods for calculating the link score, demonstrating that both measure the same concept.

As we do in Equation 1, we assume that the link score is 0 for all links from a flow to a stock if the net flow of the stock ($i - o$) is 0. As discussed above, the assumption of a 0 link score in this case does not change any loop scores since any link originating from a stock whose value does not change, will have a score of 0. We can safely set to 0 all link scores from a flow to a stock whose value does not change because link scores are multiplied into loop scores and any loop which passes through such a stock is inactive (and will have score 0).

In this formulation, a value of 0 for an inflow or outflow will result in a 0 link score. If the inflow and outflow are nearly balanced, so that there is only a small change in the stock, the scores for the links from the flows to the stock will be large, yet close in value. This happens because the denominator of Equation 2 approaches 0 faster than the numerator in such cases.

This formulation for capturing the effect of a flow on a stock is different from both the PPM and the Loop Impact method and is the characteristic of the LTM approach that allows LTM to define a single number representing the importance of a whole loop. Both the PPM and the Loop Impact method consider links involving integration in the same manner as they consider algebraic links. Specifically, they measure the change in a stock value relative to the change in value of the associated flow (the second derivative). That limits the scope of those methods so as only to address the effects of a loop on a single stock. Equation 2 directly uses the flow value relative to the change in value of the associated stock (the first derivative).

**Link score computation examples**

Table 1 demonstrates the process for calculating the link score magnitude for an auxiliary (non-stock) variable. It uses the equation $z = 2x + y$ to demonstrate how to calculate a link score magnitude. In this specific case, there are two link score magnitudes that must be calculated, one for the link x→z and one for the link y→z. To calculate the link score magnitude for the link x→z, first determine Δz which is the actual change in $z$ (5). Next determine the partial change in $z$ with respect to $x$, represented with the symbol $\Delta_x z$, by substituting into the equation for $z$ the previous value of $y$ (4) and the current value of $x$ (7). Then take the computed value of $z$ using those values (18) and subtract from it the previous value of $z$ (14) to yield $\Delta_x z$ (4). To complete the calculation of the link score magnitude, divide $\Delta_x z$ (4) by Δz (5) to get the result that 4/5ths of the change in $z$ is caused by the change in $x$.

*Table 1: Components necessary to calculate the link score magnitude for the links x→z and y→z based on the equation z = 2x+y.*

| Variable | Time 1 | Time 2 | Variable Change | Partial Change in $z$ | Link Score Magnitude |
|---|---|---|---|---|---|
| $x$ | 5 | 7 | $\Delta x = 2$ | $\Delta_x z = 4$ | $\left\lvert\frac{\Delta_x z}{\Delta z}\right\rvert = \frac{4}{5}$ |
| $y$ | 4 | 5 | $\Delta y = 1$ | $\Delta_y z = 1$ | $\left\lvert\frac{\Delta_y z}{\Delta z}\right\rvert = \frac{1}{5}$ |
| $z = 2x + y$ | 14 | 19 | $\Delta z = 5$ | - | - |

The absolute value and sign are used in the link score definitions because the change in the dependent variable may be positive or negative for reasons unrelated to the change in the independent variable. For example, in Table 2 $\Delta \mathrm{z}$ is negative while the partial change in z ($\Delta_x z$) is positive, therefore the link score magnitude before taking the absolute value ( $\frac{\Delta_x z}{\Delta \mathrm{z}}$) is negative even though $x$ has a positive influence on $\mathrm{z}$. In Table 2, incorrect polarities would result if we did not use the absolute value in computing the link score magnitude.

To measure polarity and complete the link score calculation, we must multiply the link score magnitude by the polarity ($sign\left(\frac{\Delta_x z}{\Delta \mathrm{x}}\right)$) as defined by Richardson (1995). In Table 2, we apply this to obtain the correct polarities. To calculate the polarity of the link x→z, start by following the same procedure as in Table 1 to calculate the partial change in z ($\Delta_x z$). That procedure yields a $\Delta_x z$ of 2/3 (0.67). Next determine the change in $x$ which, in this case, is 2. Finally take the sign of $\Delta_x z$ (0.67), divided by $\Delta \mathrm{x}$ (2), which is +1. This is the polarity for the link x→z.

*Table 2: Demonstration of correct polarity when calculating the link score magnitude for the links w→z, x→z, and y→z based on the equation z =* $(w + x)/y$ *.*

| Variable | Time 1 | Time 2 | Variable Change | Partial Change in $z$ | Link Score Magnitude | Link Polarity | Link Score |
|---|---|---|---|---|---|---|---|
| $w$ | 7 | 10 | $\Delta \mathrm{w} = 3$ | $\Delta_w z = 1$ | $\left\lvert\frac{\Delta_w z}{\Delta \mathrm{z}}\right\rvert = 5$ | $sign\left(\frac{\Delta_w z}{\Delta \mathrm{w}}\right) = +1$ | $\left\lvert\frac{\Delta_w z}{\Delta \mathrm{z}}\right\rvert \cdot sign\left(\frac{\Delta_w z}{\Delta \mathrm{w}}\right) = 5$ |
| $x$ | 2 | 4 | $\Delta \mathrm{x} = 2$ | $\Delta_x z = 0.67$ | $\left\lvert\frac{\Delta_x z}{\Delta \mathrm{z}}\right\rvert = 3.33$ | $sign\left(\frac{\Delta_x z}{\Delta \mathrm{x}}\right) = +1$ | $\left\lvert\frac{\Delta_x z}{\Delta \mathrm{z}}\right\rvert \cdot sign\left(\frac{\Delta_x z}{\Delta \mathrm{x}}\right) = 3.33$ |
| $y$ | 3 | 5 | $\Delta \mathrm{y} = 2$ | $\Delta_y z = -1.2$ | $\left\lvert\frac{\Delta_y z}{\Delta \mathrm{z}}\right\rvert = 6$ | $sign\left(\frac{\Delta_y z}{\Delta \mathrm{y}}\right) = -1$ | $\left\lvert\frac{\Delta_y z}{\Delta \mathrm{z}}\right\rvert \cdot sign\left(\frac{\Delta_y z}{\Delta \mathrm{y}}\right) = -6$ |
| $\boldsymbol{z} = \frac{(\boldsymbol{w} + \boldsymbol{x})}{\boldsymbol{y}}$ | 3 | 2.8 | $\Delta \mathrm{z} = -0.2$ | - | - | - | - |

**Defining loop scores**

The loop score is defined below in Equation 3 and where $L_x$ refers to the loop $x$ being studied. $LS(s_1 \to t_1)$ refers to the link score for the first link in the loop from the independent source variable $s_1$ to dependent target variable $t_1$. Loop scores are calculated by multiplying all link scores in the loop from that first one, $s_1 \to t_1$, to the last, $s_n \to t_n$, where n is the number of links in $L_x$ (so that $t_n$ is the same as $s_1$). Note that this expression multiplies both the magnitude and the sign of the different link scores, - with an odd number of negative links yielding a negative loop and an even number yielding a positive loop. The multiplication of link scores is consistent with the chain rule of differentiation as demonstrated in Appendix II and accurately represents the pacifying effect of an inactive link in an otherwise active loop. This means that any loop containing an inactive link is assigned the loop score 0.

*Equation 3: Definition of loop score, for the loop x which contains n links for source variable S to the target variables T.*

$$Loop\ Score(L_x) = \left(LS(s_1 \to t_1)\ \cdot\ LS(s_2 \to t_2)\ \dots \cdot LS(s_n \to t_n)\right)$$

The loop score, like the link score, is a dimensionless quantity. Just as the link score can be thought of as the force that one variable applies to another, the loop score can be thought of as the force one feedback loop applies to the behavior of all the stocks (and hence all the variables) it connects.

The definition of loop score is distinct from the Loop Impact of Hayward & Boswell (2014) because in the Loop Impact method the products of impacts equals the loop gain, whereas the loop score will always compute to 1 in an isolated loop as discussed in Appendix II. When comparing all loops, the loop score measures the relative importance of each feedback loop to the behavior of a model, rather than to the behavior of a single stock as is done in PPM and the Loop Impact method.

In order to establish a common baseline for comparing the contribution of feedback loops, we need to identify which feedback loops to include in the comparison. For every example presented in this paper, we analyze all feedback loops. For models where every stock in the model has a path to and from every other stock in the model, we compare the loop score across all loops in the model. For models where this is not true, we only compare the loop scores across all loops which effect the same sub-set of all the stocks in the model. The LTM method does not require that loops be independent for the analysis to be valid and therefore does not restrict analysis to the minimal independent loop set identified by Oliva (2004), or any other identified set of loops. We leave for future work to determine how to best reduce the feedback loop complexity of larger models as discussed by Kampmann (2012) and note that the work of Güneralp (2006) and Huang et. al, (2012) show that restricting loops under consideration can filter out important loops. In what follows, we consider all identified connected loops, independent or not in the topological sense.

Once the set of feedback loops being analyzed has been determined, we define the relative loop score of each loop by dividing the loop score by the sum of the loop scores of all loops in the set. Equation 4 shows this computation for the loop $X$. The summation is over feedback loops (including $X$) being analyzed. The sign of a relative loop score still represents the polarity of the feedback loop in question. The relative loop score is a normalized measure taking on a value between -1 and 1. It reports the polarity and fractional contribution of a feedback loop to the change in value of all stocks at a point in time. By comparing relative loop scores, we determine which loops contribute the most to the behavior of all stocks in the feedback loop set under study.

*Equation 4: Definition of the relative loop score for the loop X normalized over all loops n analyzed in the chosen loop set.*

$$Loop\ Score_{L_X} = \left(\frac{Loop\ Score(L_X)}{\sum_{Y=0}^{n}|Loop\ Score(L_Y)|}\right)$$

Loop scores, like link scores, can become very large and difficult to interpret without this normalization. This is especially true as an equilibrium is approached because the denominator of Equation 2 (the net flow of a stock) approaches zero faster than any individual flow. The same is true of Equation 1 where the change in the target variable, z, may approach zero regardless of the large partial change in the target with respect to the source. We demonstrate the asymptotic behavior of the loop score in the analysis of the Bass diffusion model. In that case, even though the score of the loops effectively approaches infinity, the transition from positive to negative loop dominance is smooth and clearly visible when using relative loop scores because they are normalized. An example of loop score values that can't be meaningfully compared is shown below in the case of the inventory workforce model, where a feedback loop for smoothing demand is not comparable with the others since it is not coupled with inventory-workforce adjustment process.

**Computational considerations**

Using the LTM method, we make our computations as time progresses in the model. The first computation can be made only after the model has been initialized and moved forward in time. In the results we present, we use the model's dt or time step to determine how often to compute link and loop scores. This is most straightforward using the Euler integration method. In principle the computation could proceed also at a longer or shorter sampling interval, allowing it to work with other integration methods such as Runge-Kutta.

The computational efficiency of this method has not been examined in depth and, in particular, it has not been analyzed in the case of large models with hundreds of stocks and millions of feedback loops. For models of smaller size and complexity (between 2 and 20 stocks, and less than 50 feedback loops), we have found through experience that the largest computational burden is not caused by what is required to calculate the link and loop scores, but rather by the calculations required to identify the full set of feedback loops.

An implication of the calculation method we present, is that the equations in the model will be computed not just once as it is typically done to simulate a model, but repeated once for each independent variable in the equation. This can multiply the number of computations by  2 or more (depending on equation complexity), which is similar to the computational requirements for linearization.  In short, however, the computation times are modest and quite similar to simply simulating the model.  For reference, the analysis of all of the models in the paper, including Forrester's 10 stock market growth model takes less then 1 second, including the time to parse the XMILE representation of the model, find all the loops, partition the loop set, and calculate all loop dominance metrics presented.

The computation of the link and loop score metrics, nonetheless, does require that equations be computed multiple times per dt and this cannot be done in standard software.  For this paper, we have modified an open source and publicly available simulation engine sd.js (Powers, 2019) to simulate the model and perform the link and loop score calculations. The pseudo code for this computation is shown below in Figure 1.

```
for (let target in model.variables) {
  let value = target.currentValue;
  let previousValue = target.previousValue;

  if (target.isStock) {
    let sumOfFlows = 0;
    for (let source in target.sources) {
      if (target.isInflow(source))
        sumOfFlows += source.previousValue;
      else
        sumOfFlows -= source.previousValue;
    }

    for (let source in target.sources) {
      if (sumOfFlows == 0) {
        LINKSCORE[source,target] = 0;
      } else if (target.isInflow(source)) {
          LINKSCORE[source,target] = ABS(source.previousValue / sumOfFlows);
      } else {
        LINKSCORE[source,target] = -ABS(source.previousValue / sumOfFlows);
      }
    }
  } else if (value == previousValue) {
    for (let source in target.sources) {
      LINKSCORE[source,target] = 0;
    }
  } else {
    for (let source in target.sources) {
      let tRespectSource = <calc. target, use current source, prev. of rest>;
      let deltaTRespectS = tRespectSource - previousValue;
      let deltaSource = source.currentValue - source.previousValue;
      let deltaTarget = value - previousValue;
      let sign = 1;

      if (deltaSource != 0 && deltaTRespectS!= 0) {
        sign = SIGN(deltaTRespectToS / deltaSource);
      }

      LINKSCORE[source,target] = ABS(deltaTRespectS / deltaTarget) * sign;
    }
  }
}
```
*Figure 1: Pseudo code for calculating all link scores in a model after calculating a dt of the model*

Figure 1 shows how we walk through all the variables in the model calculating the link scores for all links coming in to each variable. First we check, for each variable (*target*), whether it is a stock or not. If the *target* is a stock then we directly apply Equation 2 to determine the link score from each flow (*source*) into the target. If *target* is not a stock, we apply Equation 1. When applying Equation 1 for each *source / target* combination, the first thing we do is determine what the value for *target* would have been if only *source* changed (*tRespectSource*). To calculate *tRespectSource*, we call another subroutine which is capable of recalculating *target* using the current value of *source*, and the previous value of all other variables. Using *tRespectSource*, we determine the partial change in *target* with respect to *source* (*deltaTRespectS*) . The change in *source* and *target* are straight forward calculations using the current and previous values of *target* and *source*. Finally, we calculate the sign, making sure we do not produce a divide by zero, and then the link score, using the already calculated values.

**Application of the LTM method to the Bass diffusion model**

We have used a variant of the Bass diffusion model (Bass, 1969), pictured in Figure 2 below, to demonstrate the ability of the LTM method to reproduce the standard explanation for the behavior of that model. Richardson (1995), using his dominant polarity method, says the following about how a logistic model (e.g. the Bass diffusion model) works:

> *In the logistic equation, a shift in loop dominance occurs when the level reaches half its maximum value, the point of inflection in the logistic curve.*

This understanding has been confirmed by Kampmann and Oliva (2017) who apply traditional, manual, loop dominance analysis to reach this same conclusion.

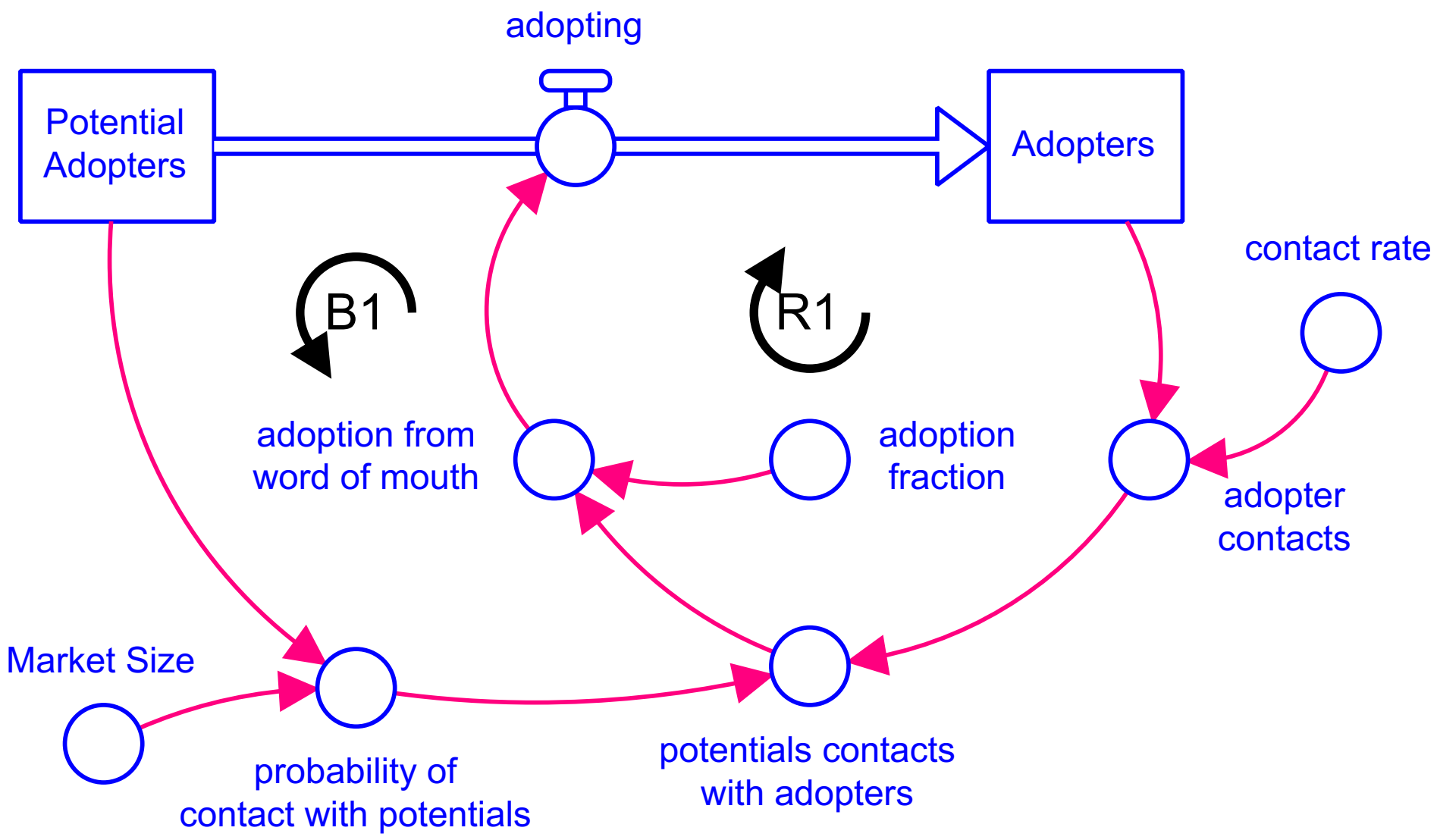

*Figure 2: The stock and flow structure of the Bass diffusion model analyzed*

This version of the Bass diffusion model runs from Time 0 to Time 15 with the inflection point reached between time 9.5625 and 9.625. The value of Market Size is 1,000,000 people initially of whom 1 are adopters . The value of the contact rate is 100 people per person per year, and the value of the adoption fraction is 0.015 (dimensionless). The equations for this model follow the standard formulation. This model contains two loops, one balancing and one reinforcing.

- Balancing (B1)
  - probability of contact with potentials
  - potentials contacts with adopters
  - adoption from word of mouth

- adopting
- potential adopters

- Reinforcing (R1)
  - adopter contacts
  - potentials contacts with adopters
  - adoptions from word of mouth
  - adopting
  - adopters

The results of the LTM analysis which appear in Table 3 and Figure 3 demonstrate that LTM reproduces the same standard explanation for behavior as Richardson (1995), Kampmann and Oliva (2017) report. In Table 3 the calculation of the loop score and relative loop score of B1 and R1 at specific points in time is demonstrated. Table 3 shows that the two loops shift in dominance during the time between time 9.5625 and 9.625 which is when the inflection occurs. In addition, Table 3 confirms that the proper polarity is assigned to each loop and link. Figure 3 supports the standard explanation of the model's behavior by showing that the relative loop score magnitude for both loops passes through 0.5, the threshold for dominance, at the inflection point.

*Table 3: Loop scores and relative loop scores in the Bass diffusion model calculated to 4 significant digits*

| Link | $T_1$ | $T_{9.5}$ | $T_{9.5625}$ | $T_{9.625}$ | $T_{15}$ |
|---|---|---|---|---|---|
| probability of contact with potentials → potentials contacts with adopters | 0.000 | 9.958 | 9358 | 10.91 | 1.000 |
| potentials contacts with adopters → adoption from word of mouth | 1.000 | 1.000 | 1.000 | 1.000 | 1.000 |
| adoption from word of mouth → adopting | 1.000 | 1.000 | 1.000 | 1.000 | 1.000 |
| adopting → potential adopters | -1.000 | -1.000 | -1.000 | -1.000 | -1.000 |
| Potential adopters → probability of contact with potentials | 1.000 | 1.000 | 1.000 | 1.000 | 1.000 |
| B1 Loop Score | 0.000 | -9.958 | -9358 | -10.91 | -1.000 |
| B1 Relative Loop Score | 0.000 | -0.465 | -0.488 | -0.512 | -1.000 |
| adopter contacts → potentials contacts with adopters | 1.000 | 11.46 | 9806 | 10.41 | 0.000 |
| Adopters → adopter contacts | 1.000 | 1.000 | 1.000 | 1.000 | 1.000 |
| adopting → Adopters potentials contacts with adopters → adoption from word of mouth | 1.000 | 1.000 | 1.000 | 1.000 | 1.000 |
| adoption from word of mouth → adopting | 1.000 | 1.000 | 1.000 | 1.000 | 1.000 |
| R1 Loop Score | 1.000 | 11.46 | 9806 | 10.41 | 0.000 |
| R1 Relative Loop Score | 1.000 | 0.535 | 0.512 | 0.488 | 0.000 |

Table 3 identifies which links contribute most critically to the feedback loop score changes. The majority of links have a link score of 1.000 because the target variables for those links only have a single variable which points to them. This is true of both the auxiliaries and the stocks. The only link score which exhibits a change in contribution over time in the loop B1 is the link 'Probability of contact with potentials → potentials contacts with adopters'. This is because that link is the only link which has a changing score. It is the key link in the loop B1 and is responsible for the changes in B1's loop score and is therefore critical to the overall shift in loop dominance. The link ('adopter contacts→ potentials contacts with adopters') is its counterpart in the loop R1 and it is just as critical to the overall shift in loop dominance, - from R1 to B1. Figure 2 shows that these two links are located at the junction between the reinforcing and balancing feedback loops. Independently, by just examining the structure, we can confirm that these links are most important with regard to the model behavior because these links points to the variable where the feedback loops interact with each other after tracing through the loops starting with the stocks.

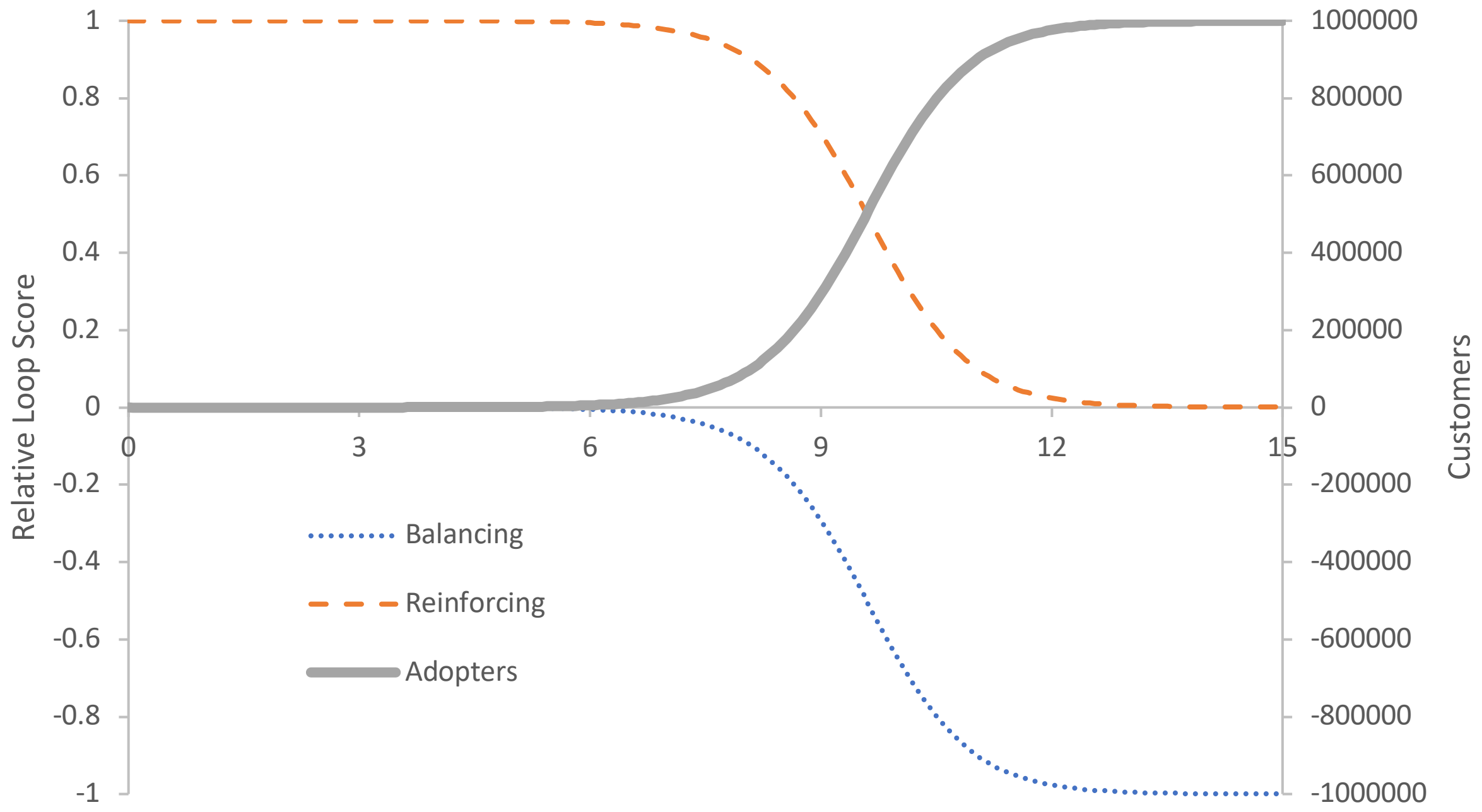

*Figure 3: Bass diffusion relative loop scores plotted over time against Adopters.*

If we were to simulate this model with a dt approaching 0, then the loop score magnitude for both loops, R1 and B1, would approach infinity at the inflection point. This happens because the Δz values from the link scores (Equation 1) for the two links into 'potentials contacts with adopters' approach 0, so that both loops have a score approaching infinity. In single stock systems, instants in time when loop scores approach infinity, represent shifts in the feedback loop dominance of the model, which in the case of the bass diffusion model is when both loops are pushing very hard to change behavior, and are, therefore, canceling each other out. This is markedly different from what we find using PPM or the Loop Impact method where the infinities occur at maximum and minimum stock values (when curvature is no longer changing). In the PPM-based approaches the zeroes represent the inflection points.

The loop score values in Figure 4 for the Bass diffusion model demonstrates why loop scores are compared in a relative fashion. When 'potentials contacts with adopters' passes through its maximum, Figure 4, which is on a log scale, shows that the absolute value of both loop scores approaches infinity. The drastic change in scale makes Figure 4 ineffective with regard to quickly and accurately determining the dominant loops in the system, but it does demonstrate the magnitude of the effort the loops are expending to change the stocks at each point in time. When the loop scores for a positive and negative loop in a model are both high, both loops are strongly contributing to behavior, working in opposite directions and cancel each other out resulting in a small change in the stocks. The non-normalized loop score metric, because of the dramatic magnitude changes, does not add much to our intuitive understanding of which loops are dominant. After being compared to the other loops, a simple and straightforward analysis can be conducted of the loop dominance profile of a model.

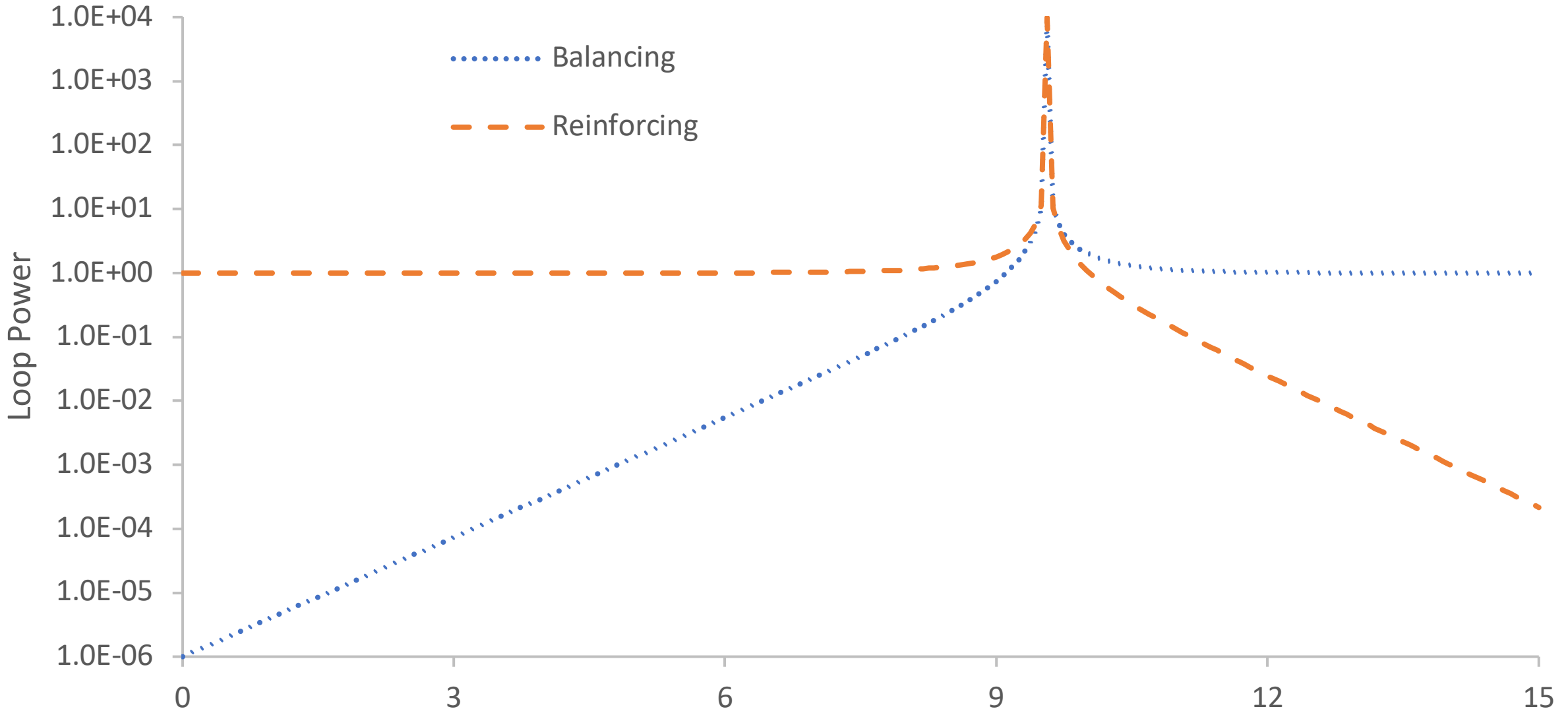

*Figure 4: The base 10 logarithm of the absolute value of Bass diffusion loop score values*

The LTM analysis of the Bass diffusion model has replicated the standard explanation for behavior in the model as performed by Richardson (1995) and Kampmann and Oliva (2017). In addition, the LTM analysis has identified the specific links in the loops that first and foremost explain the shifts in feedback loop dominance and has identified the overall effort the loops in the model are expending to change behavior.

**Application of the LTM method to the yeast alcohol model**

The yeast alcohol model is analyzed to demonstrate the efficacy of the LTM method. This analysis reinforces the notion that LTM is able to yield the same insights into behavior as previous analyses of this model using Ford's behavioral approach, PPM, Loop Impact and EEA (Saleh, 2002; Güneralp, 2005; Phaff et al., 2006; Mojtahedzadeh, 2008; Hayward and Boswell, 2014).

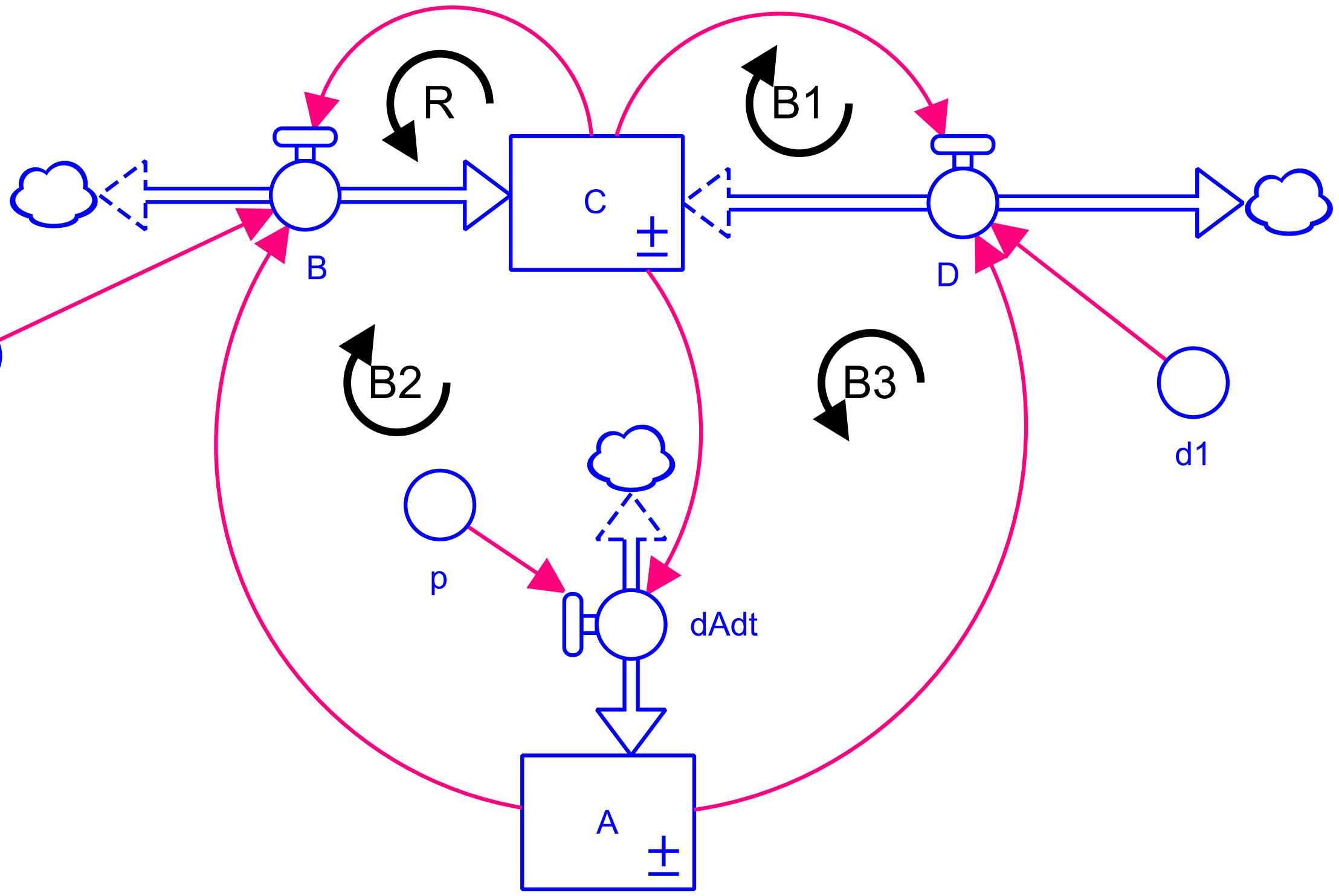

*Figure 5: The stock and flow structure of the yeast alcohol model analyzed*

Figure 5 shows the structure of the model as analyzed using a DT of .5. The model structure; B = C*(1.1-0.1*A)/b1, D=C*EXP(A-11)/d1, dAdt= p * C , is initialized with; A=0, B=1, b1=16, d1=30, and p=0.01. The model contains 4 loops, all in a single feedback loop set. Loop R, represents the birth[1] (and, late in the simulation, the deaths) of the cells C, characterized by b1 and Alcohol (A). Loop B1 represents the natural death of the cells. The main link in Loop B2 represents the slowing of the birth of cells due to the presence of alcohol. The main link in Loop B3 represents the increasing death of cells due to the presence of alcohol. This model produces the overshoot and collapse behavior seen in Figure 5 which matches exactly the behavior generated by Phaff et al. (2006) and Mojtahedzadeh (2008), and is very similar to the results that Hayward and Boswell (2014) generated using their slightly altered structure of the model[2].

[1] Notice that there is a flaw in the standard formulation of B in this model causing B to take negative values and the polarity of R to change so that it acts as an additional "deaths loop" under conditions of high levels of alcohol A. This flaw has not been corrected in order to maintain the consistency of the model across analyses in the literature.

[2] Hayward and Boswell (2014) use the same parameterization of the yeast alcohol model as us and the others, but appear to have used a Stella version of this model where uniflows were used for B and D. This subtle change to structure corrects the formulation flaw in the births loop (referenced in footnote 1) and causes their model results and loop dominance analysis to differ slightly from the other analyses and our own.

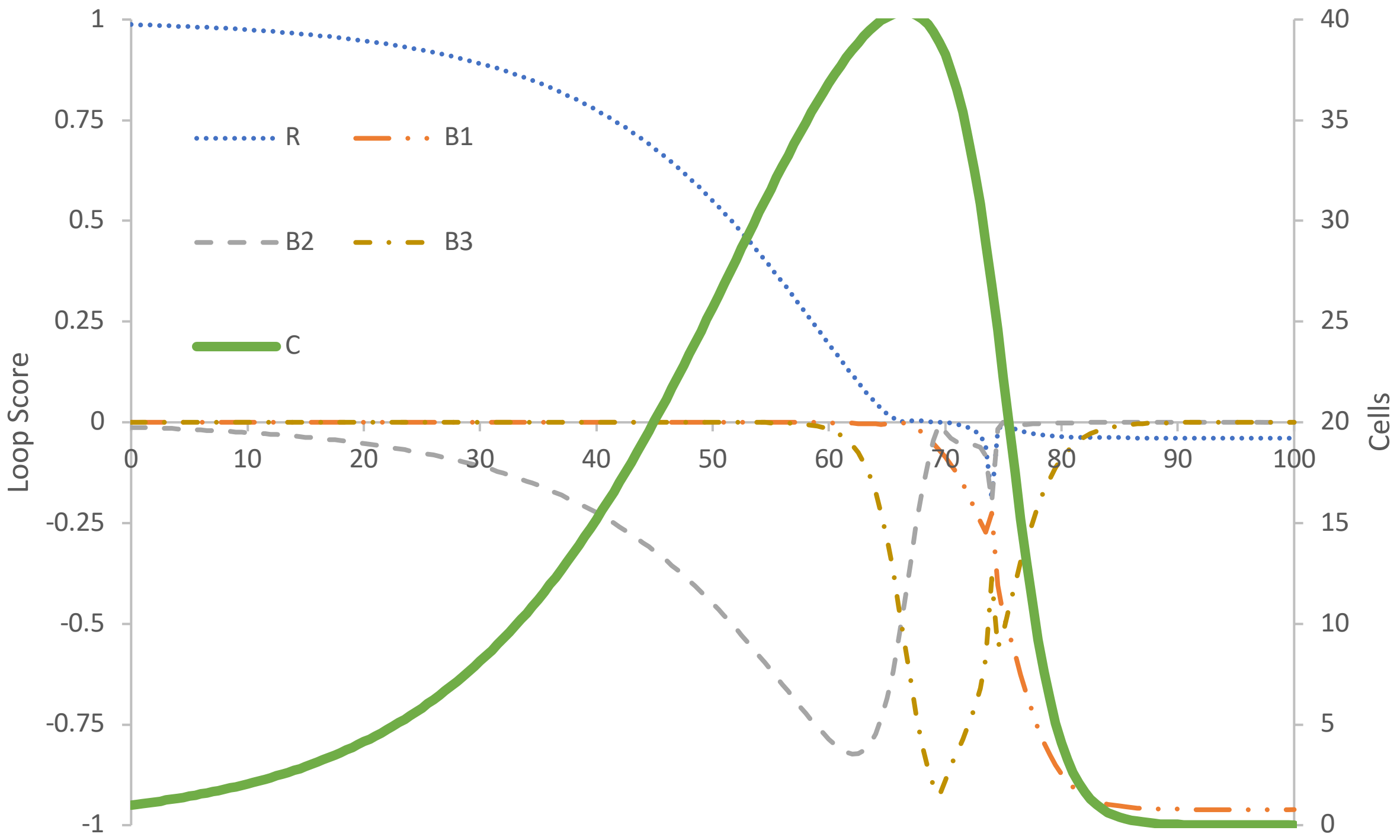

*Figure 6: Yeast Alcohol loop scores plotted against the variable C (i.e yeast cells)[3]*

*Table 4: Plot showing the relative loop scores against the variable C (yeast cells) which reveals the dominant loops in yeast alcohol model.*

| Time range | Phase 1: 0-51.5 | Phase 2: 52-66 | Phase 3: 66.5-75 | Phase 4: 75.5-100 |
|---|---|---|---|---|
| Dominant loop | R | B2 | B3[3] | B1 |

Table 4 identifies the dominant loops for each phase of the model's behavior. Comparing these results with Ford's (1999) behavioral approach as applied by Phaff et al. (2006), LTM identifies exactly the same 4 phases and agrees with the behavioral analysis in principal. LTM's only disagreement is that phase 3 is dominated by B3 alone, not B2 and B3. This same difference is raised by the PPM approach of Mojtahedzadeh (2008) and the Loop Impact approach of Hayward and Boswell (2014) where PPM and the Loop Impact method identify that Phase 3 is dominated by B3 rather than B2 and B3 together as Phaff et al.'s implementation of Ford's behavioral approach analysis would suggest. This shows that the LTM analysis of this model matches Ford's behavioral approach with the noted discrepancy and matches exactly the PPM analysis done by Mojtahedzadeh. When we compare Table 4 to Hayward and Boswell's (2014) PPM based Loop Impact method, we agree in principal with their results, but the slight change in the structure of their model[2] prevents full agreement.

---

[3] At time 74 no single feedback loop is dominant because this is the point where R is at its strongest as a balancing feedback loop (because of the model formulation error in footnote 1). After time 70 when the birth rate is negative R is acting in a similar fashion as B1. At Time 74 summing the contribution of R & B1 yields a relative loop score which is stronger than B3, but still not over 50%, B3 is the single strongest feedback loop at that exact moment and we therefore consider it alone to be dominant across phase 3.

Our results in Figure 6 and Table 4 match the EEA analysis of this model performed by Phaff et al. (2006). That analysis shows that the behavior of Phase 1 is dominated by R with B2 restraining (increasingly) the growth of C. In phase 2 EEA shows that B2 is now dominant, but R is still a significant factor in explaining the growth in C which matches what LTM concludes. This can be seen in Figure 6 because B2 has a relative loop score less then -0.5 and R is the only other active loop until time ~60 where B3 is activated in preparation for phase 3. In phase 3, EEA points to B1 and B3 together as responsible for the behavior of C. The LTM analysis matches the EEA analysis, with the caveat that the LTM analysis finds that B3[3] is solely dominant throughout that time period. EEA then reports that during phase 4, B1 is dominant over B3 which again matches the LTM analysis. That can be seen in Figure 6 as the loop score of B1 starts growing quickly at the end of phase 3, reaching nearly -1 shortly after the start of phase 4.

Looking only at the progression of relative loop scores for each of the loops as shown in Figure 6, the LTM analysis reveals some additional interesting insights. The first of these is evident from Time 74 to 76 when R becomes a significant (but not dominant) balancing feedback loop (due to the model flaw). This is due to excess levels of Alcohol (A), causing the births process to run backwards, making R into an additional loops that drains C like B1 does. At this point the relative loop scores of B2 and B3 have maximums in magnitude. This happens because those loops trade their dominance off over the course of the simulation, one rising while the other is falling. R actually has a similar local maximum in magnitude at time 74 when it reaches its peak in contribution as a balancing feedback loop. The number of stocks in a feedback loop has no direct relationship to the number, or presence, of maximums in magnitude observed in relative loop scores.

The LTM analysis of the yeast alcohol model demonstrates a shared understanding of the model with both EEA, and PPM. Moreover, the understanding coming from Ford's behavior analysis technique is also shared, demonstrating that LTM yields the same level of insight into this model as these other techniques.

**Application of the LTM method to the Inventory Workforce model, - understanding oscillations**

To demonstrate that LTM performs appropriately on a full range of SD models we have analyzed a two state oscillatory model, - a version of the inventory workforce model originally proposed by Mass and Senge's (1975). Analyzing an oscillatory system differentiates the PPM based approaches from the LTM and EEA based approaches. The version of this model analyzed is the one prepared by Gonçalves (2009) for his EEA model analysis. The model is shown in Figure 7. We use the analysis of this model also to demonstrate the impact of parameters on loop dominance patterns.

We compare and contrast the results of the LTM analysis with the ones produced by the EEA analysis Gonçalves (2009), and by two PPM-based analyses of the original Mass and Senge model by Mojtahedzadeh (2008) and Hayward and Roach (2017). The only difference between Gonçalves´version of the model (the one we analyzed) and Mass and Senge's (1975) model is the addition of the feedback loop B3 in Figure 7 which does not significantly affect the cause of the oscillation, merely the exact shape of the oscillatory mode of behavior.

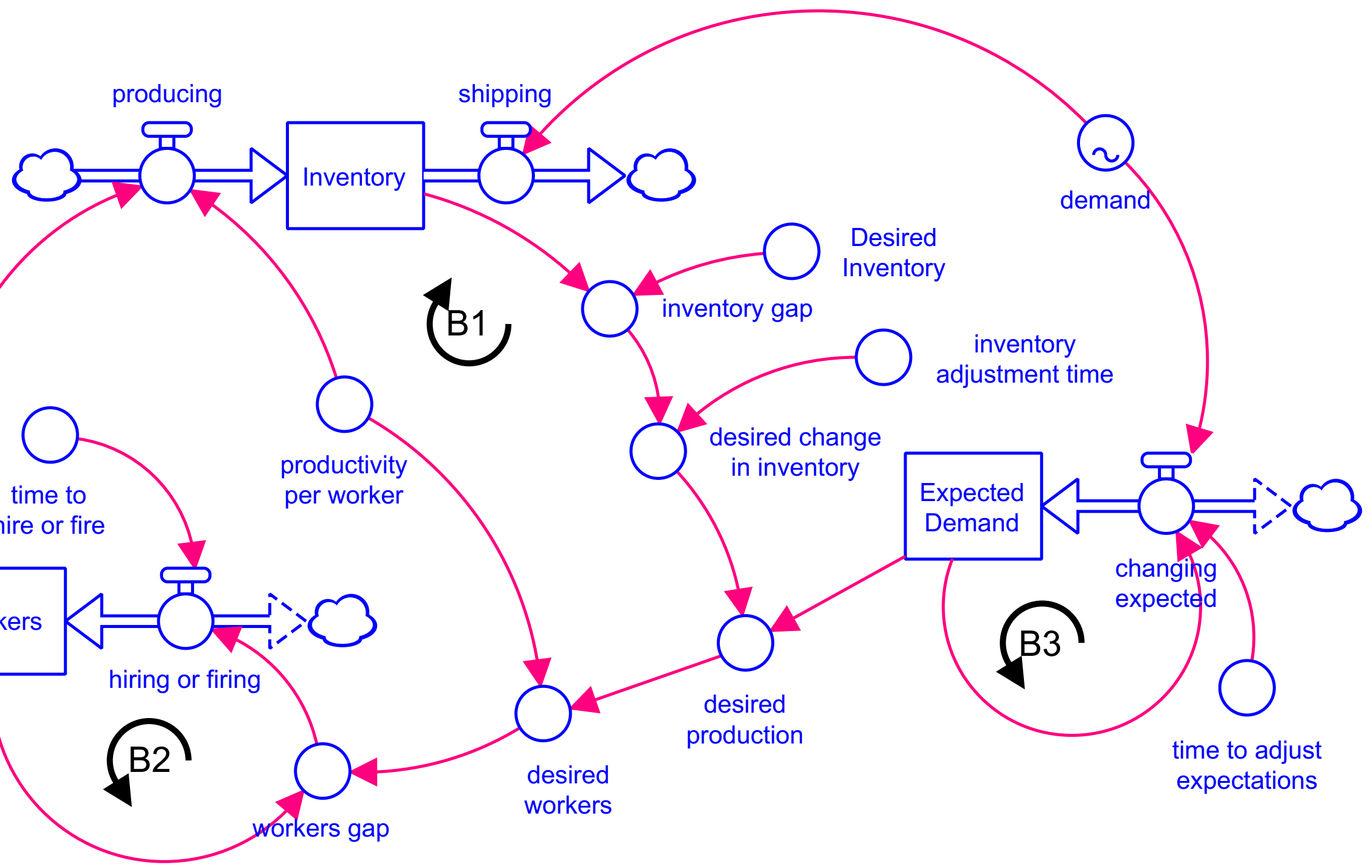

*Figure 7: The stock and flow structure of the inventory workforce model analyzed*

Our implementation of the Gonçalves(2009) model runs from Time 0 to Time 60. The model has only three balancing feedback loops that appear in two different feedback loop sets. This split happens because inventory and workforce interact, but expected demand is driven only by demand itself. The graphical function inside of the 'demand' variable acts like a step function, triggering a single increase in demand between times 1 and 2 which sets off a dampened oscillation in both workers and inventory. The loops of this model are shown in the list below, organized by loop set, containing loops that have loop scores comparable to each other, as exhibited in Figure 8.

- Loop Set 1
    - Major Balancing (B1)
        - Inventory
        - inventory gap
        - desired change in inventory
        - desired production
        - desired workers
        - workers gap
        - hiring or firing

- Workers
    - producing
  - Minor Balancing (B2)
    - Workers
    - workers gap
    - hiring or firing
- Loop Set 2
  - Expected Demand Loop (B3)
    - Expected Demand
    - changing expected

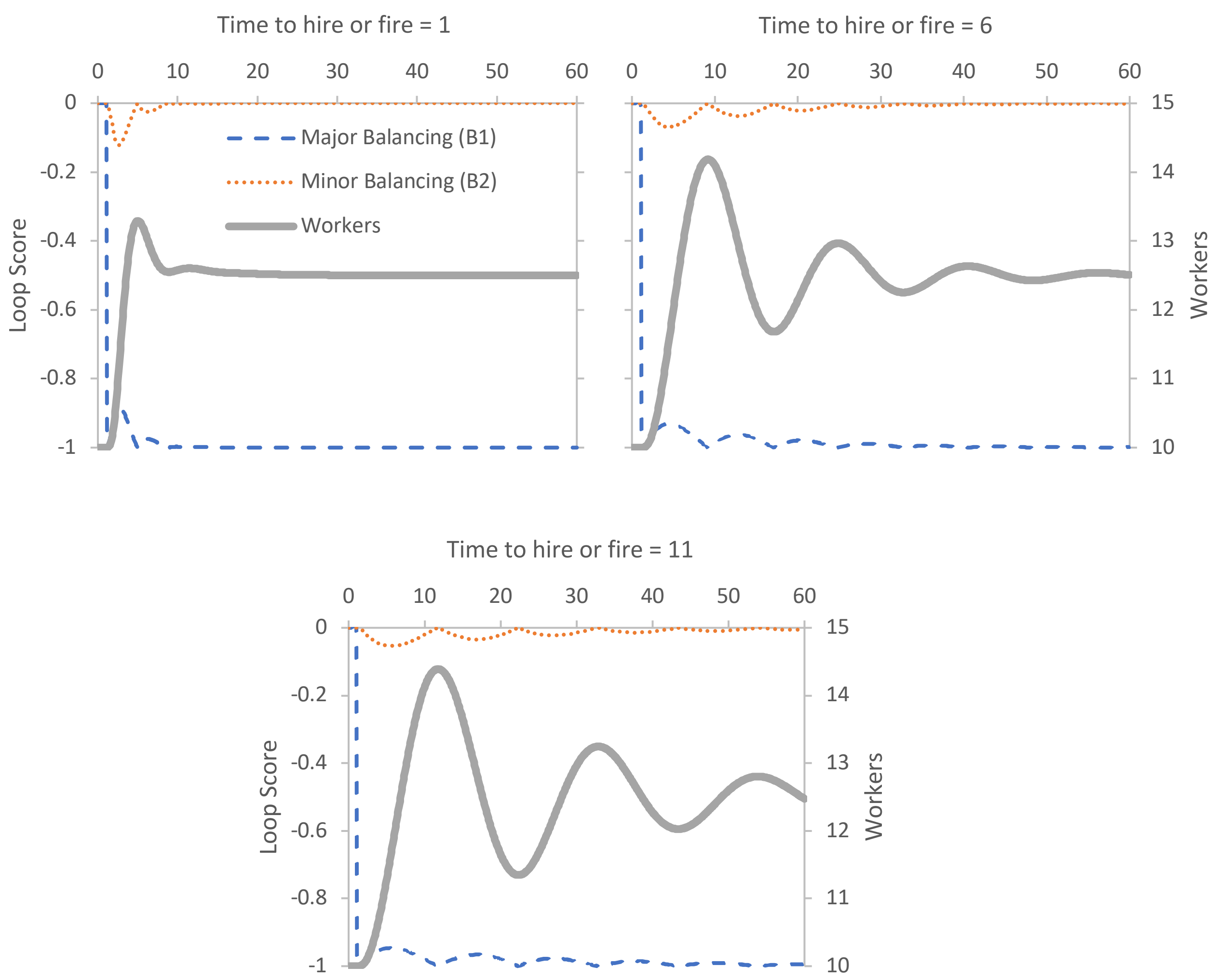

*Figure 8: Results of LTM analysis of the Inventory Workforce model showing the effect of time to hire or fire on loop dominance and Workers.*

Before the explanation of the results, note in, Figure 8, the time period before the shock in demand. During that period the model is in equilibrium, unchanging. Therefore LTM cannot inform the analysis of the model because all link scores are 0.

The two loops in this model that contain the stocks with the oscillatory behavior are B1 and B2 of Set 1. As shown in Figure 8, in all three parameterizations of the model the dominant loop describing the large majority of the change in the behavior of the worker and inventory stocks, and, therefore, the oscillations in the model is dominated by the major balancing loop B1. Figure 8 shows that there is a contribution by B2, which is dependent on the value of the parameter time to hire or fire (affecting the shape of the dampened oscillation), even though B1 is responsible for the oscillation. The longer B2 is active, the more pronounced the oscillations are. This tells us that by increasing the time to hire or fire, we increase the contribution of B2 (relative to B1). While it is true that time to hire or fire is strongly tied to the contribution of B2, it also directly impacts the contribution of B1 (i.e. independent of its impact on B2). Figure 8 shows the total relative effects of the change in this parameter on the relative contributions of B1 and B2. The net total effect of time to hire or fire on B1 and B2 causes the oscillations to become more pronounced (less damped) and to last longer.

The LTM conclusions about the impacts of time to hire or fire on the oscillation matches both the conclusions derived by EEA and PPM, - but differs from the PPM-based analyses conducted by both Mojtahedzadeh (2008) and Hayward and Roach (2017) in its explanation of dominant loops. Gonçalves (2009) EEA analysis of this model shows that the oscillatory mode of behavior arises primarily from the loop gains associated with the impacts from the major loop B1 and the damping effect is a function of the minor loop B2. The PPM-based methods show that the behavior of the model is dominated by both B1 and B2 in a cyclical process. Mojtahedzadeh (2008) states that the loop dominance pattern reported by the PPM method is not suitable for analyzing the causes of oscillation in this model, and that, instead, the explanation of the oscillation must be based on pathway frequency and stability factors which are PPM values taken at different points in a single cycle of the oscillation. Using the values of the PPM at those specific timepoints, he concludes that the source of the oscillation is B1, and that B2 is responsible for the dampening effect. This conclusion matches the results from both the EEA and the LTM analyses. The problem with PPM's shifting loop dominance pattern in oscillatory models is explained by Kampmann and Oliva (2008). They point out that methods based solely on the PPM for determining loop dominance are problematic in oscillatory in systems like the inventory workforce model (sinusoidal oscillators) because the sign of the PPM changes even though the relative contribution of the loops in the model remains constant.

The LTM analysis of the inventory workforce model has demonstrated the efficacy of the method in the analysis of oscillatory systems. The analysis of the inventory workforce model shows that the LTM method produces loop dominance patterns explaining oscillations which are the same as the EEA analyses. PPM-based analyses require more complex metrics to identify a single feedback loop as being responsible for the oscillatory mode of behavior. The LTM analysis produces conclusions about the causes of oscillation and the impacts of parameters on those oscillation that match the conclusions from both EEA- and PPM-based analyses.

**Discussion and conclusions**

As demonstrated, the LTM method provides an easy computational way to identify and understand, which feedback loops in a model dominate the model behavior or, in other words, contribute the most to the behavior of all the stocks in the model at any specific point in time. Dominance is typically model wide (with the noted exception of feedback loops that are not coupled), based on the effect on all variables, and is typically driven by the stocks that are changing most quickly in proportion to each other. As we have seen throughout the examples presented, the relative loop score measure of dominance coincides well with our structure-based understanding of relatively simple systems. The LTM method offer several important benefits.

The most important benefit of the LTM method is that it is generally applicable to all models without modification. Although, both EEA and PPM, may, in principle, be applied to all models, LTM has the advantage of being directly applicable to discrete and discontinuous models. This is because LTM analyses are conducted over time in lock step with the behavior produced by the model, do not depend on model modifications and use only values resulting from computations that are conducted as part of the normal simulation process.

The second considerable benefit of LTM is that the format of the results of the analysis are simple, easily interpretable graphs of behavior over time. LTM makes use of the existing skill-sets of all modelers and most model consumers and is thus easily accessible to all. The LTM metrics are calculated and reported over time. Therefore it takes no additional skill or training to read and interpret the graphs of the relative loop and link scores. Considerable work has been done on both PPM and EEA to define relatively easy to understand and interpret metrics. But we believe that the simplicity and accessibility of the link and loop scores provide an easier path to this type of analysis for most people.

The third noteworthy benefit to the LTM method is its relative computational and conceptual simplicity which is a major advantage over both EEA and PPM. As currently developed, the method does not use complex mathematical constructs that are not already in use by the majority of practitioners. From a mathematical perspective, the concept of the partial change in $z$ ($\Delta_x z$) is the most challenging part of the method because of its unfamiliar terminology, and not necessarily because of any inherent complexity in the idea itself. The advantage of such a simple method is that it can be understood by all practitioners so that when it comes time to apply the method, practitioners will know 'what it is saying' (i.e. how to interpret the results) due to the transparency of the method.

The fourth and final benefit of LTM is that it is relatively easily implemented in existing simulation engines without requiring modifications to existing structures within those engines. This represents a considerable advantage over EEA and PPM, and is on par with Loop Impact method.  We base this conclusion upon our own experiences implementing simulation engines in the past, - including the engines behind Stella and Vensim, and the level of effort it has taken to modify Powers' sd.js engine. This means its uptake should be relatively painless by software vendors in the field.

A weakness in the LTM method is that it is may not be used to determine loop dominance without a change in the model state.  As a model approaches equilibrium, we can see the loop scores balance one another even as they become unbounded, but when a model is in equilibrium, all loop scores are defined to be 0. Therefore, models in equilibrium cannot be analyzed using LTM.  An example of this is a simple "bathtub" population model where the birth fraction equals the death fraction.  The limitations of the link score causes the loop score for both loops to be 0 because there is no change across time.  This is a disadvantage relative to EEA which, if the model is close to linear, will provide accurate information under equilibrium conditions.   An unsatisfactory solution to this problem from a purely methodological perspective is to start introducing minute changes in these situations in order to measure those changes´ effects on loop dominance. But doing so would have major ramifications on the utility of the method for discrete and discontinuous models where information is likely encoded as specific, logically meaningful integer values.  An alternative approach would be for the model author to offset the model state from its equilibrium using a STEP function or some other modeling construct so as to expose dynamic behavior.

An additional weakness of the LTM method, which some may consider to be a strength, is that it focuses exclusively on endogenously generated behavior. Such a focus is a hallmark of System Dynamics, but is problematic for models where behavior is driven through external forcing functions that dominate the effects of feedback in the model.  The inventory workforce model contains some elements of this because of the external demand signal which necessitates the separation of the feedback loops into two sets.  But in the inventory workforce case, the external signal was only required to start the oscillation.  Loop dominance, in highly forced models (to a much greater extent than in the case of the inventory workforce model) may have little to do with behavior endogenously generated by the feedback loops. Models of this sort are currently much better analyzed using the Loop Impact method of Hayward and Boswell (2014).  Although this is one possible area of future work.

**Future work**

There are a variety of interesting extensions to LTM that combine it with other analysis techniques. The most obvious one is to combine it with Monte Carlo analysis so that the realized behavior sets encompass more of the potential behavior sets. Another is during extreme condition testing one could use LTM to show that the model is producing the right results for the right reasons.  LTM could also be combined with optimization, for example using optimizers to maximize or minimize loop scores.  This would allow practitioners to maximize the activity of favorable loops while minimizing the activity of unfavorable loops in order to automatically generate better, more robust policy recommendations.  Another area of study would include loop scores in the outputs of Monte-Carlo sensitivity analyses which would allow us to measure the robustness of loop dominance to policy or parameter changes. Monte Carlo analysis could also be used to measure the sensitivity of loop score to changes in parameter values.

A second critical area of future work is to use the LTM method to development new and exciting visualization tools, including animated stock-and-flow diagrams, where the links and flows change color and size due to changes in polarity or link score, in response to the call for such graphics by Sterman in Business Dynamics (2000). Going even further, the LTM method allows for the possibility of automated CLD generation and animation. Because the LTM method is able to tell, on a link-by-link basis, which are the key (dynamic) links in the model, it is possible, using the method, to automatically generate a CLD collapsing all of the 'unimportant' static links with scores of 0, +1.0, or -1.0 into links that are conveying a change. This will allow for an automated generation of structurally correct, minimal CLDs that accurately portray the structural components that predominantly produce the dynamics of the model, and are laid out by the computer according to best practices.

Finally, it is necessary to test and analyze larger and more varied models if we are to increase our confidence in the general utility of the LTM method. We are hopeful that the techniques laid out in this paper, will offer a significant utility and enhance the analysis and understanding of a wide set of SD models by all level of SD users.

**Appendix I: Link Scores and Partial Derivatives**

We have presented the computational equations for loop scores and link scores using separate magnitude and sign components. While this is the most intuitive presentation, it is also possible to recast the equations using partial differences which allows easier comparison with other model analysis methods.

It is straightforward to manipulate Equation 1 to have the form shown in Equation 5.

*Equation 5: Link score computed using a partial difference and magnitude adjustment.*

$$LS(x \to z) = \begin{cases} \left(\frac{\Delta_x z}{\Delta \mathrm{x}} \cdot \left|\frac{\Delta \mathrm{x}}{\Delta \mathrm{z}}\right|\right), & \\ 0, & \Delta \mathrm{z} = 0 \text{ or } \Delta \mathrm{x} = 0 \end{cases}$$

Here the first term is just the partial difference, which has the direction of change and the sensitivity of $z$ to $x$. The second term adjusts this by the realized changes so that the link score reflects the amount of contribution as opposed to the sensitivity. Dividing both the top and bottom of the second term by Δt gives Equation 6.

*Equation 6: Dividing numerator and denominator of second term by Δt*

$$LS(x \to z) = \begin{cases} \left(\frac{\Delta_x z}{\Delta \mathrm{x}} \cdot \left|\frac{\Delta \mathrm{x}/\Delta \mathrm{t}}{\Delta \mathrm{z}/\Delta \mathrm{t}}\right|\right), & \\ 0, & \Delta \mathrm{z} = 0 \text{ or } \Delta \mathrm{x} = 0 \end{cases}$$

If the model being analyzed (including $f(x, y)$) is continuously differentiable, then as Δt approaches 0, so too will Δx and we can write Equation 6 as:

*Equation 7: The simplified continuous representation of the link score metric*

$$LS(x \to z) = \begin{cases} \left(\frac{\partial z}{\partial \mathrm{x}} \cdot \left|\frac{\dot{\mathrm{x}}}{\dot{\mathrm{z}}}\right|\right), & \\ 0, & \dot{\mathrm{z}} = 0 \text{ or } \dot{\mathrm{x}} = 0 \end{cases}$$

Equation 7 demonstrates the relationship of the link score to the partial derivative which are key to Pathway Participation Metrics and the Loop Impact, which also form the basis for the linearized representation of model equations used to compute eigenvalues. The addition of the second term converts the potential contribution to the realized contribution. This is why the Loops that Matter approach gives results that are largely in line with other approaches when normalized, but at the same time are not identical in absolute value

**Appendix II: Important Analytic Characteristics of Link Scores**

There are two notable characteristics of link scores that it is useful to elucidate. The first is invariance under formulation, something that is relied upon when computing loop scores. The second is the observation that isolated loops will always have a loop score of 1 or -1.

Consider invariance under formulation changes. Put simply, it should not matter whether we connect two variables with one complicated equation, or three variables with two simpler equations. To demonstrate this, without loss of generality, suppose that there are two formulations for the variable $z$. The first is directly as $f(w, x, y)$, the second indirectly as $g(u, y)$ where $u = h(w, x)$. In this case we first compute the link score from $x$ to $u$ as shown in Equation 8, then the link score from $u$ to $z$ as shown in Equation 9.

*Equation 8: The link score from $x \to u$ in a compound formulation.*

$$LS(x \to u) = \begin{cases} \left(\left|\frac{\Delta_x u}{\Delta \mathrm{u}}\right| \cdot sign\left(\frac{\Delta_x u}{\Delta \mathrm{x}}\right)\right), & \\ 0, & \Delta \mathrm{u} = 0 \text{ or } \Delta \mathrm{x} = 0 \end{cases}$$

*Equation 9: The link score from $u \to z$ in a compound formulation*

$$LS(u \to z) = \begin{cases} \left(\left|\frac{\Delta_u z}{\Delta \mathrm{z}}\right| \cdot sign\left(\frac{\Delta_u z}{\Delta \mathrm{u}}\right)\right), & \\ 0, & \Delta \mathrm{z} = 0 \text{ or } \Delta \mathrm{u} = 0 \end{cases}$$

The composite link score (a link score multiplied along a causal pathway) is computed as the product of the two link scores as shown in Equation 8 and Equation 9, which can be rewritten as Equation 10 by multiplying by $\left|\frac{\Delta x}{\Delta x}\right|$ which is just 1.

*Equation 10: The composite link score from $x \rightarrow z$.*

$$LS(x \rightarrow z) = \begin{cases} \left(\left|\frac{\Delta_x u}{\Delta u}\right| \cdot \left|\frac{\Delta_u z}{\Delta z}\right| \cdot sign\left(\frac{\Delta_x u}{\Delta x}\right) \cdot sign\left(\frac{\Delta_u z}{\Delta u}\right)\right), \\ 0, \qquad \Delta z = 0 \text{ or } \Delta u = 0 \text{ or } \Delta x = 0 \end{cases}$$

*Equation 11: The composite link score written as chained difference equations.*

$$LS(x \rightarrow z) = \begin{cases} \left(\left|\frac{\Delta_x u}{\Delta x}\right| \cdot \left|\frac{\Delta_u z}{\Delta u}\right| \cdot \left|\frac{\Delta x}{\Delta z}\right| \cdot sign\left(\frac{\Delta_x u}{\Delta x}\right) \cdot sign\left(\frac{\Delta_u z}{\Delta u}\right)\right), \\ 0, \qquad \Delta z = 0 \text{ or } \Delta u = 0 \text{ or } \Delta x = 0 \end{cases}$$

Applying the chain rule for partial differences this is the same Equation 11. Cancelling the $\Delta x$ terms we get the expression in Equation 12.

*Equation 12: The composite link score reduced by the chain rul*

$$LS(x \rightarrow z) = \begin{cases} \left(\left|\frac{\Delta_x z}{\Delta x}\right| \cdot \left|\frac{\Delta x}{\Delta z}\right| \cdot sign\left(\frac{\Delta_x u}{\Delta x}\right) \cdot sign\left(\frac{\Delta_u z}{\Delta u}\right)\right), \\ 0, \qquad \Delta z = 0 \text{ or } \Delta u = 0 \text{ or } \Delta x = 0 \end{cases}$$

*Equation 13: The composite link score fully reduced*

$$LS(x \rightarrow z) = \begin{cases} \left(\left|\frac{\Delta_x z}{\Delta z}\right| \cdot sign\left(\frac{\Delta_x u}{\Delta x}\right) \cdot sign\left(\frac{\Delta_u z}{\Delta u}\right)\right), \\ 0, \qquad \Delta z = 0 \text{ or } \Delta u = 0 \text{ or } \Delta x = 0 \end{cases}$$

The $sign$ term in Equation 12 and Equation 13 should be clear, since it just tracks the directional change, and thus we have the same result we would get with all the computation in a single equation.

This chaining is the same as that observed in Richardson (1995) and Kampmann (2012) and allows us to look at models as they are constructed, and not in any condensed form.

It is worth noting that this equivalence fails if $\Delta u = 0$, even when both $\Delta x$ and $\Delta z$ are nonzero. That is, if the intermediate variable is not changing, the link score becomes 0 even when the input and ultimate output are changing. It is easy to construct models that have this characteristic (in fact the Bass Diffusion model can be written with total population computed by adding the two stocks), and the 0 value is helpful at showing that the potential feedback is not real.

The second observation on loop scores is that for a single positive or negative loop the score will be +/-1, though the gain around the loop could be significantly different. This is easy to see in a model of (net) population growth, since the link score from the stock to the flow will be 1 as only the stock is changing the flow, and with only a single net flow the link score from the flow to the stock will also be one. Similar logic applies to an exponential drain, though in this case the link score from the flow to the stock is -1 so the loop score becomes -1.

This value of 1 is true regardless of the population growth rate, or residence time, in the above example. This is an important distinction between the loop score and gain around a loop. It is also interesting to think about what happens to the single loop score as additional loops are added. For example, adding deaths to a population model with only births would give link scores from the flows into the stock based on their relative value. In this case both of the loop scores will have magnitude greater than 1, and the closer the flows are the bigger the scores. This is, as discussed, why the relative loop scores are reported as the basis for analysis and emphasizes how distinct the loop score is from a representation of gain. Put another way, loop scores do not predict the speed of change, but only show which part of structure is dominant at any point in time.